\documentclass[useAMS,usenatbib]{mn2e}
\usepackage[totalwidth=500pt, totalheight=680pt]{geometry}

\usepackage{epsfig,graphicx,amsmath,amssymb}
\usepackage[normalem]{ulem}

\newcommand{\hm}{h^{-1}}

\title[The origin of peak-offsets in weak-lensing maps]{The origin of
  peak-offsets in weak-lensing maps}
\author[J. P. Dietrich et al.]{J. P. Dietrich$^{1}$\thanks{E-mail:
jorgd@umich.edu (JPD)},
A. B\"ohnert$^{2}$,
M. Lombardi$^{3}$,
S. Hilbert$^{4,5}$, and
J. Hartlap$^{4}$\\
$^{1}$Physics Department and Michigan Center for Theoretical Physics,
University of Michigan, 450 Church St, Ann Arbor, MI 48109, USA\\
$^{2}$ESO, Karl-Schwarzschild-Str. 2, 85748 Garching b. M\"unchen,
Germany\\
$^{3}$Dipartimento di Fisica, Universit\`a degli Studi di Milano, via
Celoria 16, 20133 Milano, Italy\\ 
$^{4}$Argelander-Institut f\"ur Astronomie, Auf dem H\"ugel 71, 53121 Bonn,
Germany \\
$^{5}$Max-Planck-Institut f{\"u}r Astrophysik, Karl-Schwarzschild-Stra{\ss}e
1, 85741 Garching, Germany}
\begin{document}

\date{Accepted 2011 October 12. Received 2011 October 11; in original form 2011 March
21}

\pagerange{\pageref{firstpage}--\pageref{lastpage}} \pubyear{2011}

\maketitle

\label{firstpage}

\begin{abstract}
Centroid positions of peaks identified in weak lensing mass maps often
show offsets with respect to other means of identifying halo centres,
like position of the brightest cluster galaxy or X-ray emission
centroid.
Here we study the effect of projected large-scale structure (LSS),
smoothing of mass maps, and shape noise on the weak lensing peak
positions. Additionally we compare the offsets in mass maps to those
found in parametric model fits.
Using ray-tracing simulations through the Millennium Run $N$-body
simulation, we find that projected LSS does not alter the weak-lensing
peak position within the limits of our simulations' spatial
resolution, which exceeds the typical resolution of weak lensing maps.
We conclude that projected LSS, although a major contaminant for
weak-lensing mass estimates, is not a source of confusion for
identifying halo centres. The typically reported offsets in the
literature are caused by a combination of shape noise and smoothing
alone. This is true for centroid positions derived both from mass maps
and model fits.
\end{abstract}


\section{Introduction}
\label{sec:introduction}
Weak gravitational lensing is one of the most important tools to study
clusters of galaxies. Its sensitivity to both luminous and dark matter
is a distinct advantage over other methods. Observations of a
clusters' tidal shear field imprinted on the observed ellipticities of
faint background galaxies allow for ``parameter-free'' mass
reconstructions (see e.g.
\citealp{1993ApJ...404..441K,2001A&A...374..740S} for methods, and
\citealp{2001A&A...379..384C,2002A&A...395..385C,2009A&A...499..669D,2010A&A...520A..58I}
for applications). Such two-dimensional maps often show offsets
between the position of the observed peaks in surface mass density and
other indications of a galaxy cluster's centre, like the position of
the brightest cluster galaxy or the centroid of the X-ray emission
\citep[e.g.,][]{2005A&A...440..453D,2006ApJ...643..128W,2010PASJ...62..811O}.
Such offsets can be of either astrophysical origin or caused by random
or systematic noise. An example of the latter could be missing data
due to bright stars or other masked image defects.

Two examples of clusters where the observed separation between
weak-lensing and X-ray centroid has been attributed to astrophysical
causes are the ``Bullet Cluster'' \citep{2006ApJ...648L.109C} and
MACS~J0025.4$-$1222 \citep{2008ApJ...687..959B}. These are systems in
which high-speed collisions separated the collisionless dark-matter
and galaxy components from the collisional intra-cluster medium.

In other cases the spatial discrepancy has been determined to be
consistent with the centroid shifts induced by the shape noise of the
background galaxies \citep[e.g.,][]{2006A&A...451..395C}. It has also
been noted that the observed centroid offsets depend on the halo
concentration, with more centrally concentrated halos showing smaller
offsets in convergence maps \citep{2010ApJ...719.1408F}.

Weak lensing is susceptible to the entire mass along the line-of-sight
and it is now well established that projected large-scale structure
(LSS) and halo triaxiality are significant contributions to the total
error budget in weak-lensing mass estimates of galaxy clusters
\citep{2001A&A...370..743H,2003MNRAS.339.1155H,2011MNRAS.412.2095H,2004PhRvD..70b3008D,2010arXiv1011.1681B,2007MNRAS.380..149C}.
In addition to influencing weak-lensing mass-estimates, irregular halo
shapes and projected LSS could potentially shift the centroids of the
observed surface mass density distribution away from the true halo
centre.  We will summarily refer to these effects as projected LSS
throughout this work. E.g., \citet{2005ApJ...623...42L} speculated
that the projection of groups close to the line-of-sight might be the
cause for the $8\arcsec$ or $94\,\hm$\,kpc offset they
observed between the weak lensing and optical/X-ray centroids of
RCDS~$1259.9-2927$. Our aim in this paper is to study the importance
of this effect and its relative importance to centroid shifts induced
by shape noise and the inevitable smoothing in mass reconstructions.
Knowledge of all these effects is required to make robust statements
about the nature and astrophysical significance of observed peak
offsets.

The outline of the paper is as follows. We describe the simulations
and how we identified weak lensing peaks in them in
Sect.~\ref{sec:methods}. We present the results of matching halo
centres to lensing peak positions in Sect.~\ref{sec:results}. Finally,
we summarise and discuss our findings in
Sect.~\ref{sec:summary-discussion}. We use standard lensing notation
throughout \citep[e.g.][]{2006sassfeebookS}.

\section{Methods}
\label{sec:methods}

To study the spatial offsets between halo centre and lensing peak
position, we simulated weak-lensing observations of galaxy clusters
using $N$-body and ray-tracing simulations. We used the Millennium Run
\citep[MR,][]{2005Natur.435..629S} simulation. The MR is a large dark
matter $N$-body simulation with $2160^3$ particles in a
$500\,\hm$\,Mpc box simulating structure formation from $z=127$ to the
present epoch in a flat $\Lambda$CDM Universe. The assumed
cosmological parameters are: a matter density
$\Omega_\mathrm{m}=0.25$, a cosmological constant with energy density
$\Omega_\Lambda=0.75$, a Hubble constant of $h=0.73$ (in units of
$100\,\mathrm{km}\,\mathrm{s}^{-1}\mathrm{Mpc}^{-1}$), and an initial
density power spectrum normalisation $\sigma_8=0.9$. The Plummer
equivalent smoothing scale in the MR is $5\,\hm$\,kpc. 

Cluster halos in the MR were identified by a friend-of-friends
algorithm. Cluster centres were identified with the position of the
minimum of the gravitational potential. Cluster masses were obtained
by measuring the mass inside spheres around the cluster centre with
mean density 200 times the critical cosmic density.

The ray-tracing algorithm described in \citet{2009A&A...499...31H} was
used to calculate the apparent positions of the cluster centres and
the convergence in 512 $4\times4\,\text{deg}^{2}$ fields of view. In
these simulations, the matter distribution is created by periodic
continuations of the simulation boxes at increasing redshifts to
create backward light cones. Matter is divided into slices of
thickness $\approx 100\,\hm$\,Mpc perpendicular to the line-of-sight
and angled with respect to the box sides to avoid the repetition of
structures along the line-of-sight. Light rays are shot from redshift
zero to $z = 3.06$ through these lens planes. The solid angle
subtended by these simulations is small enough to neglect the effects
of sky curvature.

The products of these ray-tracing simulations are noise-free maps of
the dimensionless surface mass-density $\kappa$ and complex shear
$\gamma$. This noise-free case allows us to study the impact of LSS
projections alone, without being affected by smoothing and shape
noise, which are always present in mass reconstructions derived from
astronomical observations.

\subsection{Mass maps}
\label{sec:mass-maps}

Already \citet{1993ApJ...404..441K} noted that a mass reconstruction
obtained from an unfiltered shear field would have infinite variance.
Smoothing the shear field with a low-pass filter suppresses the noise
but also amounts to a smoothing of the reconstructed $\kappa$ field.
For a Gaussian filter $\propto \exp(-\theta^2 / 2
\sigma_\mathrm{s}^2)$ the variance in $\kappa$ is
\citep{1993ApJ...404..441K}
\begin{equation}
  \label{eq:1}
  \sigma^2_\kappa = \frac{\sigma^2_\epsilon}{16 \pi n \sigma_\mathrm{s}^2}\;,
\end{equation}
where $\sigma_\epsilon$ is the intrinsic 2-d ellipticity dispersion
and $n$ the number density of background galaxies. This smoothing also
leads to correlated noise in the mass reconstruction. The noise
power spectrum is given by
\citep{1998A&A...335....1L} 
\begin{equation}
  \label{eq:2}
  P(k) = \frac{\sigma_\epsilon^2}{4 \pi^2 n} \exp\left(-\sigma_\mathrm{s}
    k^2\right) \;.
\end{equation}
The noise on large fields is rotation and translation
invariant, and therefore the covariance between modes at different $k$
is zero. Maps of correlated noise were created by drawing independent
complex Gaussian random variables with variance given by
eq.~(\ref{eq:1}) and applying an inverse Fourier transform to the
arrays created in this way.

In this work we used three values for the number density of background
galaxies $n=\{10, 30, 80\}\,\mathrm{arcmin}^{-2}$ and two smoothing
scales $\sigma_\mathrm{s} = \{45\arcsec, 90\arcsec\}$. The two smaller
number densities correspond roughly to typical ground-based
observations with $2$\,m to $8$\,m class telescopes
\citep[e.g.,][]{2007A&A...470..821D,2010PASJ...62..811O}, while the
higher number density represents space-based data
\citep[e.g.,][]{2006ApJ...648L.109C}. The large smoothing scale was
used for the lowest number density, while the small smoothing scale
was applied in the case of the two larger number densities.

\begin{figure}
  \resizebox{\hsize}{!}{\includegraphics[clip]{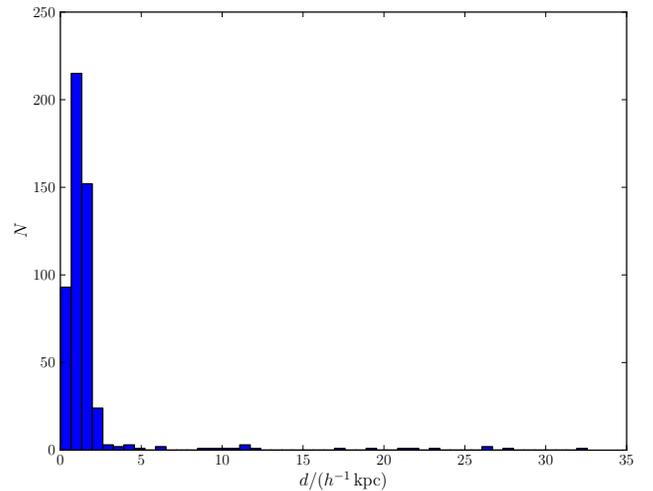}}
  \caption{Physical offsets between $512$ halo and lensing
    peak positions for a single background redshift population at $z =
    1.08$.}
  \label{fig_phys_off_z1.0}  
\end{figure}

Peaks in $\kappa$-maps were detected with a connected-component
labelling of pixels above a detection threshold. We used the
8-connectivity, i.e., we consider all pixels that are connected via
the sides, edges, or corners of a square as one
structure. Additionally, we imposed the condition that each peak must
have at least 3 pixels above the detection threshold. The pixel with
the highest value is considered to be the peak location.  We compared
this peak location to the one derived by SExtractor's
\citep{1996A&AS..117..393B} fit of a bivariate Gaussian. The two
methods give results that generally agree to within a fraction of
pixel. This deviation is too small to be a concern for us. Halos were
then matched to their nearest neighbour in the list of detected
peaks. All halos with a mass of more than $10^{14}\,\hm\,M_\odot$ and
$z < 1.1$ were considered in this matching.

\subsection{Shear catalogues and model fits}
\label{sec:shear-catalogs}

Often weak-lensing studies of clusters include parametric model fits
to the tangential shear profiles
\citep[e.g.,][]{2009A&A...499..669D,2010PASJ...62..811O}, in which the
centroid position can be treated as free parameters
\citep[e.g.,][]{2010MNRAS.405.2215O}. Since no smoothing is applied to
the shear field, shape noise and LSS projections are the only source
of noise. To compare with the centroid positions obtained from mass
maps, we created mock catalogues of galaxies with number densities
$n=\{10, 30, 80\}$\,arcmin$^{-2}$ and intrinsic ellipticities drawn
from a Gaussian with 2-dimensional ellipticity dispersion
$\sigma_\epsilon = 0.38$, truncated at $|\epsilon| = 1$.  The reduced
shear $g$, computed from the ray-tracing, was applied to these
intrinsic ellipticities using the standard relation
\citep{1997A&A...318..687S}
\begin{equation}
  \label{eq:3}
    \epsilon =
  \begin{cases}
    \frac{\epsilon^{\mathrm{(s)}} + g}{1 + g^*\epsilon}  & |g|
    \le 1 \; , \\
    \frac{1 -
      g\epsilon^{*\mathrm{(s)}}}{\epsilon^{*\mathrm{(s)}} +
      g^*} & |g| > 1 \; ,
  \end{cases}
\end{equation}
where the asterisk denotes complex conjugation, and the superscript
$(s)$ denotes intrinsic source properties.

We fitted NFW \citep{1997ApJ...490..493N} models to the shear
estimated from the galaxy catalogue within a radius of $15\arcmin$
around the halo centre positions. We fixed the NFW concentration
parameter to follow the \citet{2004A&A...416..853D} mass-concentration
relation, leaving three free parameter, the halo centre
$(x_\mathrm{c}, y_\mathrm{c})$ and the halo mass $m_\textrm{200}$.
Fits were performed using the maximum likelihood estimator of
\citet{2000A&A...353...41S}.

\section{Results}
\label{sec:results}

\subsection{Noise-free case}
\label{sec:noise-free-case}

Figure~\ref{fig_phys_off_z1.0} shows the apparent and physical offsets
between centres of halos and associated peak positions in noise free
$\kappa$-maps. While the matching between halos and lensing peaks was
performed for all cluster sized halos at redshifts below $1.1$ as
described in the previous section, we made a redshift dependent mass
cut in the selection of halos that enter our comparison. We selected a
fiducial minimum cluster mass of $M_\mathrm{fid} = 2\times
10^{14}\,\hm\,M_\odot$ at redshift $z_\mathrm{fid} = 0.3$. The minimum
mass is governed by the lensing efficiency and scales with the halo
redshift $z$ as
\begin{equation}
  \label{eq:4}
  M_\mathrm{min}(z) = M_\mathrm{fid} 
  \frac{\mathcal{D}(0, z_\mathrm{fid}) \mathcal{D}(z_\mathrm{fid},
  z_\mathrm{bgk})}{\mathcal{D}(0, z) \mathcal{D}(z, z_\mathrm{bgk})} \;,
\end{equation}
where $\mathcal{D}(\cdot, \cdot)$ is the angular diameter distance
between two redshifts, and $z_\mathrm{bgk}$ is the redshift of the
background source population. Here the background sources were assumed
to be at a single redshift $z_\mathrm{bgk} = 1.08$, which is
  the redshift of the first lens plane at $z > 1$.

Obviously, almost all lensing peaks are projected to the immediate
vicinity of the halo centre. Virtually no associations are made on
scales $> 15\,\hm\,\mathrm{kpc}$, which are beyond the resolution
limit of the Millennium Run simulation. Visual inspection of the few
large-distance matches confirmed that their large offsets are not
caused by LSS projections. They are undetected halos erroneously
matched to a nearby lensing peak or their surface mass density signal
was merged into a double peaked structure when it overlapped with a
more significant foreground or background object. Our peak finder does
not attempt to deblend such objects into multiple peaks, although a
human observer would correctly identify the smaller peak as caused by
the halo. Thus, we do not consider these cases to be LSS projections
for the purpose of this work. Furthermore, in optical multi-colour
data, both clusters would be readily identifiable as separate
entities.

While a source redshift of $z \sim 1$ is typical for the mean redshift
of targeted ground-based lensing observations, the tail of the
redshift distribution extends to higher values, and space based
observations have the potential to probe LSS at significantly higher
redshifts. We therefore matched the same set of halos to convergence
peaks detected in maps with a single source plane at $z = 3.06$, the
highest redshift in our ray-tracing simulations, in an effort to
maximise the impact of LSS projections.

The result is virtually unchanged. The $90\%$ percentile increases
from $2.0\,\hm$\,kpc to $5.6\,\hm$kpc and a few more halos have barely
resolved offsets, but still 478 out of the 512 halos have offsets of
less than $15\,\hm$\,kpc, down from 503 in the case of the lower
source redshift.

\begin{figure}
  \resizebox{\hsize}{!}{\includegraphics[clip]{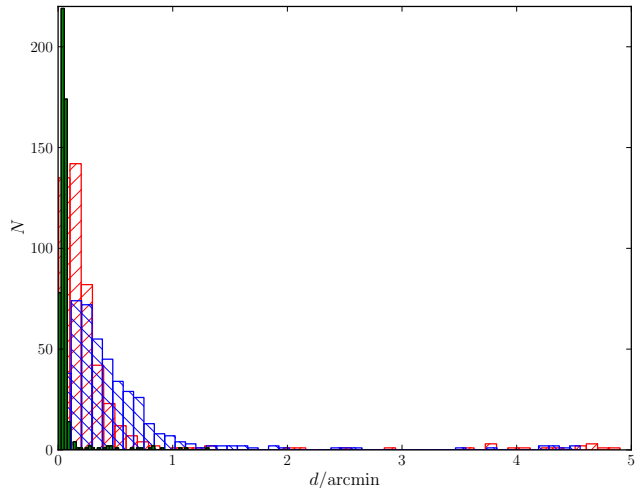}}
  \caption{Apparent offsets in smoothed $\kappa$-maps. The red/blue
    (rising/falling hatches) histograms are for smoothing scales of
    $45\arcsec$ and $90\arcsec$, respectively. The thin solid green
    histogram is the unsmoothed case from Fig.~\ref{fig_phys_off_z1.0},
    shown here for comparison.}
  \label{fig:smoothed_offsets}
\end{figure}

\subsection{Smoothing and shape noise}
\label{sec:smooth-shape-noise}

We now turn to the influences of smoothing and shape noise. The effect
of smoothing is shown in Fig.~\ref{fig:smoothed_offsets}.  $474$
($434$) halos fulfilling our selection criterion in eq.~(\ref{eq:4})
could be matched to peaks detected in $\kappa$-maps smoothed with
Gaussians of width $45\arcsec$ ($90\arcsec$) within a distance of
$5\arcmin$. Again, the vast majority of large separation pairs in the
tail of the distribution is due to erroneous associations. Ignoring
outliers at $d > 2\arcmin$, we find that almost 90\% of all
associations are made within one half of the smoothing scale. The
observed offsets are, however, significantly larger than in the
absence of smoothing, as is shown by the comparison with the
unsmoothed case in Fig.~\ref{fig:smoothed_offsets}.  Thus, smoothing
alone, without the added influence of shape noise, causes shifts of
centroid positions with respect to the halo centres.

We summarise properties of the offset distributions in
Table~\ref{tab:kappa-offsets}. Here the mode of the distributions is
estimated from the maximum of a Gaussian kernel density estimate of
the offset distribution. It is worth pointing out that for one-sided
distributions whose mode is very close to zero, this method has a
systematic bias towards higher values. Nevertheless, the values we
find are generally in good agreement with naive mode estimates
obtained from looking at the histogram plots, with the exception of
the LSS only case. All other quantities are computed directly from the
measured offsets.

\begin{figure}
  \resizebox{\hsize}{!}{\includegraphics[clip]{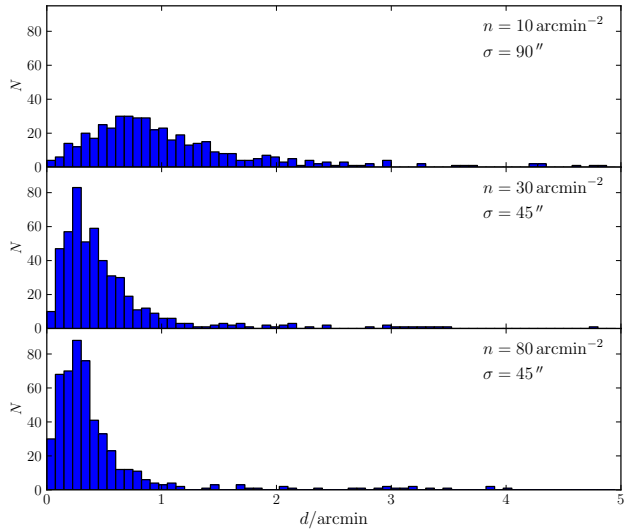}}
  \caption{Apparent offsets in smoothed noisy $\kappa$-maps. The three
  panels are for different smoothing scales and number densities as
  denoted in their legends.}
  \label{fig:noise_offsets}
\end{figure}

Finally, Fig.~\ref{fig:noise_offsets} shows the observed offsets in
smoothed maps with shape noise added. As described in
Sect.~\ref{sec:methods}, we assumed combinations of number densities
and smoothing scales of $(n/\mathrm{arcmin}^{-2},
\sigma_\mathrm{s}/\arcsec = \{(10, 90), (30, 45), (80, 45)\}$. As one
would expect, the apparent offsets are larger than in the noise free
case. The distribution's mode roughly triples for $n =
10\,\mathrm{arcmin}^{-2}$ and approximately doubles for $n = \{30,
80\}$\,arcmin$^{-2}$. The distributions also have noticeably
longer tails than in the noise free case. Again, the distribution
properties are summarised in Table~\ref{tab:kappa-offsets}.

\subsection{NFW model fits}
\label{sec:nfw-model-fits}

\begin{figure}
    \resizebox{\hsize}{!}{\includegraphics[clip]{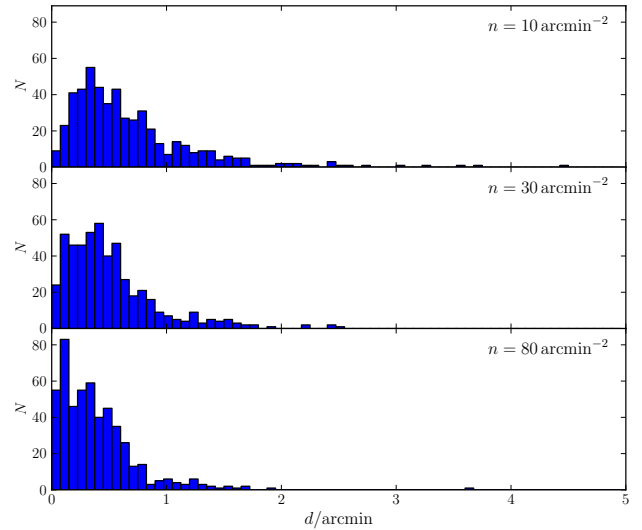}}
  \caption{Apparent centroid offsets in NFW model fits for three different number densities.}
  \label{fig:nfw-offsets}
\end{figure}

The centroid offset distributions obtained from fitting NFW models to
the shear catalogues, as described in
section~\ref{sec:shear-catalogs}, are shown in
Fig.~\ref{fig:nfw-offsets} and the distribution properties are
summarised in Table~\ref{tab:kappa-offsets}. The offsets are smaller
than those observed in $\kappa$-maps for the low number density case
only. For $n=\{30, 80\}$\,arcmin$^{-2}$ the offsets are comparable, if
not slightly worse than in noisy mass maps.

\begin{table*}
  \caption{Properties of the offsets distributions in different
    $\kappa$-maps and in the model fits as a function of number
    density $n$ and smoothing scale $\sigma_\mathrm{s}$.}
  \begin{tabular}{rrrrrrr}
    \hline\hline
    $n/\mathrm{arcmin}^{-2}$ & $\sigma_\mathrm{s}/\arcsec$ & mode/$\arcsec$ &
    mean/$\arcsec$ & $90\%\mathrm{ile}/\arcsec$ &
                                $95\%\mathrm{tile}/\arcsec$ \\\hline
    \multicolumn{6}{c}{LSS only}\\
    $\infty$ &  0 &  9 &  4 &  5 &   9 \\\hline
    \multicolumn{6}{c}{Smoothed maps}\\
    $\infty$ & 45 &  8 & 13 & 26 &  33 \\
    $\infty$ & 90 & 14 & 25 & 48 &  59 \\\hline
    \multicolumn{6}{c}{Smoothing and shape noise}\\
    10       & 90 & 44 & 52 & 87 & 101 \\
    30       & 45 & 17 & 27 & 52 &  64 \\
    80       & 45 & 15 & 21 & 41 &  51 \\\hline
    \multicolumn{6}{c}{NFW model fits}\\
    10       & 0 & 22 & 37 & 72 & 85 \\
    30       & 0 & 22 & 29 & 58 & 76 \\
    80       & 0 &  9 & 23 & 45 & 62 \\\hline
    \hline
  \end{tabular}
  \label{tab:kappa-offsets}
\end{table*}

\section{Summary and Discussion}
\label{sec:summary-discussion}

We used ray-tracing simulations through the Millennium Run $N$-body
simulation to study the impact of projected large-scale structure on
the observed peak positions of halos in weak-lensing maps. The main
result of this work is that such projections do not lead to shifts of
the maximum surface mass density away from the halo centre. This is
true even for the extreme case of a single source redshift at $z = 3$.
Thus, this is a robust result applicable to both ground- and deep
space-based lensing observations. By making the mass selection in
eq.~(\ref{eq:4}) dependent only on the lensing efficiency in our mock
survey, we neglected selection biases that will effect real galaxy
cluster catalogues. Real clusters are often studied in detail
precisely because they reveal striking lensing features, offsets
between dark and luminous matter, or other unusual features.
Furthermore, mergers, projections, and alignment of triaxial halos
with the line-of-sight will inevitably boost low mass halos to appear
to be above the mass selection. While this could in principle be a
concern for the applicability of our results to a highly biased
cluster sample, the virtually complete absence of significant offsets
caused by LSS projections across our entire sample suggests that this
is not a concern.

The negligible effect that LSS projections have is exemplified by the
case of RCDS~$1259.9-2927$ at $z = 1.24$, which we mentioned in the
introduction. Its lensing centroid has an offset of $8\arcsec$.  This
corresponds to a projected distance of $48\,\hm$\,kpc and is
significantly larger than any of the offsets seen in
Fig.~\ref{fig_phys_off_z1.0}. Thus, this offset is most likely not
caused by projected LSS. We emphasise that, although
Table~\ref{tab:kappa-offsets} suggests that an offset of $8\arcsec$ is
consistent with LSS projections at the $\sim 95\%$ confidence level,
we see such large offsets only for lower redshift halos. The relevant
comparison here is with the physical offsets reported in
Fig.~\ref{fig_phys_off_z1.0}. This is supported by looking at the
offsets caused by LSS as a function of redshift. We grouped the 512
halos into four equally large groups of increasing redshifts and
report the median and 95th percentile of their respective offset
distributions in Table~\ref{tab:z-dependence}. The median shows a
slight increase with redshift but still well below the resolution of
our simulations. The tails of the distributions show no clear trend.
We conclude that the offsets caused by LSS are limited in physical
scales to those scales that are much smaller than the redshift
independent smoothing scale for all halos studied in this work.

\begin{table}
  \caption{Offsets caused by LSS projections as a function of halo redshift.}
  \begin{tabular}{rrrr}\hline\hline
    $z_\mathrm{min}$ & $z_\mathrm{max}$ & 
    median/$\hm$\,kpc & 95\%tile/$\hm$\,kpc\\\hline
    0.07 & 0.34 & 1.1 & 10.5 \\
    0.34 & 0.46 & 1.6 &  6.6 \\
    0.46 & 0.58 & 1.8 &  3.4 \\
    0.58 & 0.93 & 2.1 &  5.6 \\
    \hline\hline
  \end{tabular}
  \label{tab:z-dependence}
\end{table}
Smoothing an asymmetric mass distribution with a symmetric kernel
naturally leads to centroid shifts. We showed that these occur on
scales less or equal to one half of the smoothing scale.  Adding shape
noise leads to additional, significantly larger offsets.

For typical shallow ground-based observations with number
density $10\,\mathrm{arcmin}^{-2}$ and smoothing scale $90\arcsec$,
the mode of the offset distribution is at $\sim 44\arcsec$ and the
mean of the distribution is at $52\arcsec$.  These values are
compatible with those found by blind searches for galaxy clusters
\citep{2007A&A...462..459G,2007A&A...462..875S,2010ApJ...709..832G}.
All these values were obtained by taking only those halos into account
which could be matched to a weak-lensing peak within $5\arcmin$.
Including the long tail of matches at larger separations of course
leads to the inclusion of mostly spurious matches. The $75\%$
percentile of the full distribution is at $2\farcm3$, reproducing the
result of \citet{2007A&A...470..821D}.

Obviously the offsets become smaller if a higher number density of
background galaxies suppresses the shape noise. However, even with the
depths achieved by current space-based data sets, the offsets are
significantly larger than in the absence of shape noise.  E.g., for
$n=80$\,arcmin$^{-2}$ the mode of the offset distribution is at
$15\arcsec$, almost twice as large as in the noise-free case where it
is at $8\arcsec$. The situation is slightly better for the tail of the
distribution where the 95th percentile is still larger by a factor of
$1.5$.

Our findings show that the significance of peak offsets can often be
reliably estimated by bootstrapping from the shear catalogue, as it
was e.g. done by \citet{2006ApJ...648L.109C}. Such a bootstrapping
procedure does not include the systematic centroid shifts due to
smoothing but this effect is often smaller than the offsets typically
caused by shape noise. The resulting underestimation of offset
distances can be rectified by taking into account that smoothing leads
to additional offsets of up to half the smoothing scale at 90\%
confidence.  Applying this to the example of the Bullet Cluster, we
conclude that the contribution of smoothing is absolutely negligible
and the bootstrapping procedure provided a reliable estimate of the
offsets' significances.

\citet{2010MNRAS.405.2215O} studied the offsets of the weak lensing
signal of 25 massive clusters in the LoCuSS sample from their
brightest cluster galaxies (BCG). When comparing their results to our
NFW model fits, it is important to remember that they have no
knowledge of the true halo centre as we have in the MR but must
instead rely on the BCG position as an estimate for the halo centre.
This is subject to possible misidentifications of the BCG or
displacements of the BCG with respect to the dark matter halo centre,
as discussed in detail in \citet{2007arXiv0709.1159J}. However,
\citeauthor{2010MNRAS.405.2215O} conclude that their result is
consistent with a $10\%$ fraction of clusters with BCG
misidenitifications, in agreement with the work of
\citet{2007arXiv0709.1159J}. The remaining offsets are all smaller
than $1\arcmin$. This is comparable to what we find in the middle
panel of Fig.~\ref{fig:nfw-offsets}, which has approximately the same
number density as the LoCuSS data set. However, our results also
suggest that $\sim 10\%$ of all offsets are larger than $1\arcmin$.
This could very well masquerade as BCG misidentifications, which
should occur with the same frequency.

\section*{Acknowledgements}
This work was initiated during a visit by JPD to ESO {Gar\-ching}. JPD
would like to the thank the ESO visitor programme for support and the
ESO staff for hospitality. JPD gratefully acknowledges support from
NSF grant AST 0807304. We thank an anonymous referee for a number of
comments and question that helped to improve this paper significantly,
as well as Eduardo Rozo and Tim McKay for helpful comments on the
draft version of this paper. SH and JH acknowledge support by the
Deutsche Forschungsgemeinschaft within the Priority Programme 1177
under the project SCHN 342/6 and the Transregional Collaborative
Research Centre TRR 33 ``The Dark Universe''.

\bibliographystyle{mn2e}
\bibliography{offsets}

\bsp
\label{lastpage}

\end{document}